\documentclass{elsart}

\usepackage{graphicx}
\usepackage{amssymb}

\begin{document}

\begin{frontmatter}



\title{
Staggered grids discretization in three-dimensional Darcy convection
%Three-dimensional Darcy convection and cosymmetric family of steady states
}

\author{
Karas\"{o}zen B.\thanksref{label2}
%\corauthref{cor1}
}
\ead{bulent@metu.edu.tr}
\author{Nemtsev A.~D.\thanksref{label3}}
,
\author{Tsybulin V.~G.\thanksref{label3}}
%\corauth[cor1]{bulent@metu.edu.tr}
%\ead{tsybulin@math.rsu.ru}
\address[label2]{Department of Mathematics and Institute of Applied Mathematics,
\\
Middle East Technical University, Ankara, Turkey}
\address[label3]{Department of Computational Mathematics,
\\
Southern Federal University, Rostov-on-Don, Russia}

%\author{
% \footnote{E-mail:bulent@metu.edu.tr}
%\\
%
%\\
% \footnote{E-mail:tsybulin@math.rsu.ru}
%\\
%}
%
%\address{}

%    General info
%\subjclass[2000]{Primary 73Q65, 65M06; Secondary 35B20, 35K60, 90A16}
%\date{January 1, 1994 and, in revised form, June 22, 1994.}

\begin{abstract}
We consider three-dimensional convection of an incompressible fluid
saturated in a parallelepiped with a porous medium. A mimetic
finite-difference scheme for the Darcy convection problem in the
primitive variables is developed. It consists of staggered
nonuniform grids with five types of nodes, differencing and
averaging operators on a two-nodes stencil.
The nonlinear terms are approximated using special schemes.
Two problems with different boundary conditions are considered to
study scenarios of instability of the state of rest.
Branching off of a continuous family of steady states was detected
for the problem with zero heat fluxes on two opposite lateral
planes.
\end{abstract}

\end{frontmatter}

\maketitle

%\end{document}

\section*{Introduction}
Several algorithms ( finite element, finite difference, finite volume and spectral methods) are used  for the simulation of physical problems involving partial differential equations (PDEs).
Usually, the PDEs express fundamental physical laws like conservation  of mass, momentum and total energy.
While solving such problems, some information about the problem and
its structure is lost during the discretization that replaces the
PDE by a system of algebraic equations.
In recent years so-called mimetic discretizations  were developed to yield stable and accurate solutions by preserving the analytical properties of the underlying PDEs \cite{HymMorShaSte02,BocHym06}.
Several mimetic or conservative finite difference methods were derived and their conservation properties were discussed for PDEs arising in fluid dynamics \cite{MorLunVasMoi98,Vas00,HamLieStr02}.
Experience shows that discrete conservation of mass, momentum and kinetic energy produce better results as compared with nonconservative methods.

%
%Many algorithms are used for the simulation of physical problems
%involving partial differential equations (PDEs). Among these
%algorithms, the finite element, finite difference, finite volume and
%spectral methods are prominent. Usually, the PDEs express
%fundamental physical laws like conservation  of mass, momentum and
%total energy.
%While solving such problems, some information about the problem and
%its structure is lost during the discretization that replaces the
%PDE by a system of algebraic equations.
%In recent years many methods were developed to preserve many of  the
%properties of the continuous problem by the discretization as
%possible. Among these methods  are the so-called mimetic
%discretizations for PDEs which yield stable, accurate approximate
%solutions by preserving the analytical properties of the underlying
%PDEs \cite{HymMorShaSte02,BocHym06}. Several mimetic or conservative
%finite difference methods were developed in recent years for PDEs
%arising in fluid dynamics. Experience shows that discrete
%conservation of mass, momentum and kinetic energy produce better
%results as compared with nonconservative methods. In
%\cite{MorLunVasMoi98,Vas00,HamLieStr02} several conservative finite
%difference schemes on uniform and staggered grids were derived and
%their conservation properties were discussed.

The main goal of this paper is to develop a mimetic scheme for three-dimensional equations of the convection in a porous medium.
Natural convection of incompressible fluid in a porous medium
differs from the convection of a single fluid \cite{NiBe99}.
Usually, after the state of rest loses its stability, a finite
number of regimes (convective patterns) may appear.
An exciting example with an
infinite number of steady states was found for the planar problem of
incompressible fluid convection in a porous medium based on the
Darcy law \cite{Lyu75}. This case of the appearance of a continuous
family of equilibria was explained by the  cosymmetry theory
\cite{Yud91,Yud95}.

To compute  a continuous family of steady states we need the numerical
scheme to be mimetic of the underlying system. The approximation of
the planar Darcy convection equations was done for the first time
using the Galerkin method \cite{Gov98}. Then the computations of the
families of steady states  were performed using the
finite-difference scheme \cite{KarTsy99,KarTsy05a} and a combination
of spectral and finite-difference approaches \cite{KanTsy02}. In
\cite{KarTsy05} a staggered grid discretization for the planar
problem of Darcy convection  was developed in primitive variables.
Special approximation of the nonlinear terms of the underlying
system was the crucial step   in all these works. It was found that
the loss of gyroscopic and cosymmetric properties in the
finite-dimensional approximation destroys the family of steady
states and leads to a finite number of isolated stationary regimes
\cite{KarTsy99,KanTsy02}.

In this work we consider the three-dimensional problem of natural
convection in a porous medium and develop a finite-difference scheme
for a system in primitive variables (velocities, pressure and
temperature). The discretization is based on nonuniform staggered
grids \cite{KarTsy05} with five types of nodes by using the
differencing and averaging operators on two-nodes stencil. The
present finite difference scheme is constructed in the spirit of the
fully  conservative second-order finite difference scheme for
incompressible flow on nonuniform grids \cite{HamLieStr02}. An
algorithm for the computation of the family of steady states is
described, and is  used in numerical computations.

This paper is organized as follows.
The equations for the three-dimensional Darcy convection problem in primitive variables are formulated in Section 1.
In Section 2, the grids, discrete operators and discretization in space are described.
%NEW
The computation of the continuous family of steady states with
varying stability spectra  (cosymmetric family) is described in
Section 3. Numerical results  on the branching off of the family of
steady states and isolated regimes from the state of rest are
presented in Section 4.
%NEW

\section{Darcy convection}
\subsection{Darcy equations in primitive variables }
We consider an enclosure filled by a porous medium saturated by an incompressible fluid which is heated from below.
%NEW
We assume that the fluid is incompressible according to the
Boussinesq approximation \cite{NiBe99}.
%NEW
 The system of equations consists of the momentum equation based on the Darcy law
\begin{equation}
\label{eq_vec}
\frac{1}{\varepsilon} \dot{ v} =
-\nabla p + \theta \gamma - v,
\end{equation}
and the continuity and energy equations
\begin{equation}
 \label{eq_con}
 \nabla \cdot v =0,
\end{equation}
\begin{equation}
 \label{eq_tem}
  \dot \theta =
  \Delta \theta - \lambda v \cdot \gamma - (v \cdot \nabla) \theta.
\end{equation}
Here $v=(v^1,v^2,v^3)^T$ is the  velocity vector,
$\gamma=(0,0,-1)^T$ is the direction of gravity,
$p(x,y,z,t)$ is the pressure,
$\theta(x,y,z,t)$ is the deviation of the temperature from a linear (in $z$) profile, $\epsilon$ is the porosity of the
medium,
$x$, $y$, $z$ are the space variables and
the dot accent  denotes differentiation w.r.t. time $t$.

The Rayleigh number is defined as $\lambda = \alpha g T K l /\chi \mu$,
where $\alpha$ is the thermal expansion coefficient,
$g$ is the gravity acceleration,
$\mu$ is the kinematic viscosity,
$\chi$ is the thermal diffusivity of the fluid,
$K$ is the permeability coefficient,
$T$ is the characteristic temperature difference
and  $l$ is the length parameter.
 We suppose that the temperature at the boundary is given by a  linear function on the vertical coordinate $z$.

The parallelepiped $\mathcal D = [0,L_x] \times [0,L_y] \times [0,L_z]$, with length $L_x$, depth $L_y$ and height $L_z$ is filled with the fluid.
The normal component of the velocity is equal to zero at the boundary
\begin{equation}
\label{bc_vel}
V \cdot n = 0 , \quad (x,y,z) \in \partial \mathcal D.
\end{equation}

We consider  two problems with different  boundary conditions for
the temperature. The first problem is characterized by a temperature
deviation $\theta$ equal to zero at the boundary $\partial \mathcal
D\/$ (Dirichlet boundary condition):
\begin{equation}
\label{bc_A}
\theta = 0,  \quad  \quad (x,y,z) \in \partial \mathcal D.
\end{equation}

The second problem has mixed boundary conditions: the heat flux equals zero on two lateral faces $\partial_1 D = \{y=0\} \cup \{y=L_y\}$ and the temperature deviation $\theta$ is equal to zero on the remaining faces $\partial_2 D = D \setminus \partial_1 D$
\begin{equation}
\label{bc_B}
\theta_y = 0 ,  \quad (x,y,z) \in \partial_1 D ,
\quad
\theta = 0  ,  \quad (x,y,z) \in \partial_2 D .
\end{equation}

The initial condition is given as follows
\begin{equation} \label{ic}
\theta(x,y,z,0) = \theta_0(x,y,z),
\quad v(x,y,z,0) = v_0(x,y,z) .
\end{equation}
%because the other unknowns can be found as solutions of  system
%(\ref{eq_vec}), (\ref{eq_con}), (\ref{bc_vel}) and (\ref{bc_A}) (or (\ref{bc_B})).

It is simple to check that the equations
(\ref{eq_vec})--(\ref{bc_A}) are invariant with respect to the
discrete symmetries
\begin{eqnarray}
\label{sym_x}
&&
R_x  : \{x,y,z, v^1,v^2,v^3,p,\theta \} \mapsto
\{L_x -x,y,z, -v^1,v^2,v^3,p,\theta \},
\\
\label{sym_y}
&&
R_y  : \{x,y,z, v^1,v^2,v^3,p,\theta \} \mapsto
\{x,L_y -y,z, v^1,-v^2,v^3,p,\theta \},
\\
\label{sym_z}
&&
R_z  : \{x,y,z, v^1,v^2,v^3,p,\theta \} \mapsto
\{x,y,L_z -z, v^1,v^2,-v^3,p,-\theta \}.
\end{eqnarray}
%NEW
This implies that with appropriate transformations of velocity,
pressure and temperature deviation, a given set of solutions
produces a new set.
%NEW

\subsection{Darcy equations for the temperature and stream function}
When the initial velocity $v_0$ and the initial temperature distribution $\theta_0$ do not  depend on $y$ the system (\ref{eq_vec})--(\ref{eq_tem}), (\ref{bc_A}) has a  two-dimensional solution
\begin{equation}
\nonumber
v^1 = v^1(x,z,t), \;
v^2=0, \;
v^3 = v^3(x,z,t), \;
p = p(x,z,t), \;
\theta = \theta(x,z,t).
\end{equation}
%\begin{eqnarray*}
%\label{2d_sol}
%v^1 &=& v^1(x,z,t), \quad
%v^2=0, \quad
%v^3 = v^3(x,z,t),
%\\
%p &=& p(x,z,t), \quad
%\theta = \theta(x,z,t)
%, \quad v_0 = v_0(x,z)
%.
%\end{eqnarray*}
In this case we can write our system as a system containing
temperature  and stream function.
We follow the usual assumption
in porous media flow and neglect inertia in the momentum equation.
The continuity equation (\ref{eq_con}) is fulfilled automatically
when the stream function $\psi$ is given by
\begin{equation} \label{psi}
 v^1 = - \psi_z, \quad v^3 = \psi_x.
\end{equation}
Then the underlying system can be transformed  to a new form. After
application of the $curl$-operator to (\ref{eq_vec}) we deduce
\begin{equation} \label{eq_psi}
 0 = \Delta_2 \psi - \theta_x \equiv G , \quad
 \Delta_2() = ()_{xx} + ()_{zz},
\end{equation}
and using (\ref{psi}) we obtain from (\ref{eq_tem})
\begin{equation}   \label{eq_theta}
   \dot \theta = \Delta_2
\theta + \lambda \psi_x + J(\theta,\psi) \equiv F ,
\quad
 J(\theta,\psi) = \theta_x \psi_z - \theta_z \psi_x.
\end{equation}
The boundary conditions for the system (\ref{eq_psi}),
(\ref{eq_theta}) follow from (\ref{bc_B}):
\begin{equation} \label{bc_psi}
  \psi = \theta = 0 , \quad (x,z) \in \partial
  \widehat{\mathcal D}, \mbox{ where } \widehat{\mathcal D} =[0,L_x] \times [0,L_z] .
\end{equation}
and the initial condition is formulated only for the temperature
\begin{equation}
\label{ic_psi}
\theta (x,z,0) = \theta_0 (x,z),
\end{equation}
where $\theta_0$ denotes the initial temperature distribution. For a
given $\theta_0$, the stream function $\psi$ can be obtained from
(\ref{eq_psi}) and (\ref{bc_B}) as the solution of the Dirichlet
problem via Green's operator $\psi = G\theta_x$.

Equations (\ref{eq_psi})--(\ref{bc_psi}) require that  the equilibrium $\theta=\psi=0$ (state of rest),
be stable if $\lambda < \lambda_{11}$, where $\lambda_{nm} = (2 \pi n /a)^2 + (2
\pi m /b)^2$ ($m, n \in Z$) are the eigenvalues for the
corresponding spectral problem.
 It was shown in \cite{Yud95} that the first critical
value $\lambda_{11}$ has  multiplicity two for any domain $D$. As a
result, a continuous family of steady states appears
\cite{Lyu75,Yud91}.
%Then whenever  $\lambda$ passes through $\lambda_{nm}$ ($n+m>2$), this produces a new family with unstable equilibria.

The system (\ref{eq_psi})--(\ref{bc_psi}) possesses the cosymmetry
property \cite{Yud91}: a vector-function $(\theta, -\psi)$  being
orthogonal to the right-hand side of (\ref{eq_psi}) and
(\ref{eq_theta}) in $L_2$.
 Then, we obtain  the cosymmetry condition in the following form
\begin{equation}
\label{int}
 \int_{ \widehat{\mathcal D}\/} (F \psi - G \theta) dx dz
 = \int_{ \widehat{\mathcal D}\/}
\left[
\Delta \theta \psi - \Delta \psi \theta + \lambda \psi_x \psi +
\theta_x \theta + J(\theta,\psi) \psi
\right] dx dz  = 0.
\end{equation}
 This can be checked directly using integration by parts and Green's formulae.

For the  integration of  equations (\ref{eq_psi})--(\ref{bc_psi}) it is essential to provide a  discrete version of the cosymmetry condition.
 In \cite{KarTsy99} a regular uniform mesh was used and both stream function and temperature were defined at the same nodes.
The Jacobian approximation was based on the Arakawa scheme \cite{Ara66}  and a number of one-parameter families of steady states were computed.
 It was also found  that a violation of the cosymmetry
property led to a degeneration of the family.
 The application of staggered nonuniform grids for the problem (\ref{eq_psi})--(\ref{ic_psi}) was considered  in \cite{KarTsy05}.

\section{Spatial discretization}
\begin{figure}[h]
%\begin{center}
\includegraphics[width=15.0cm]{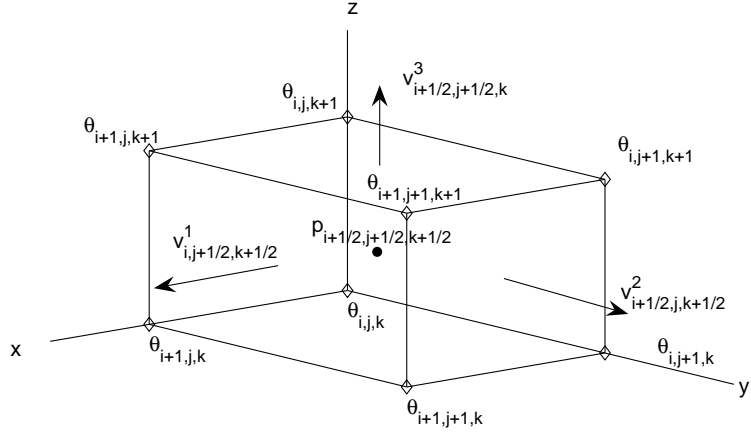}
\caption{Grid and nodes}
%\end{center}
\label{grid_nodes}
\end{figure}

We have discretized the equations (\ref{eq_vec})--(\ref{ic}) using
five different types of nodes: one for the pressure,  another for
the temperature  and three nodes for the components of the velocity
vector, see Fig.~\ref{grid_nodes}.

For the first problem with Dirichlet boundary conditions (\ref{bc_vel}) and  (\ref{bc_A}) we introduce a nonuniform grid for the temperature $\theta$
\begin{eqnarray*}
%\label{tsyb-grid}
&&
0=x_0<x_1< \ldots <x_{N_x+1}=L_x ,
\\
 \nonumber
 &&
0=y_0<y_1< \ldots <y_{N_y+1}= L_y ,
\\
 \nonumber
 &&
0=z_0<z_1< \ldots <z_{N_z+1}=L_z .
\end{eqnarray*}
The grid for the second problem with mixed boundary conditions (\ref{bc_vel}) and   (\ref{bc_B}) differs only in $y$ direction
$$
y_0+y_1=0, \quad y_1< \ldots <y_{N_y}, \quad y_{N_y}+y_{N_y+1}=2 L_y .
\nonumber
$$
Thus the temperature $\theta$ is defined at the nodes
$$
 \omega_0 = \{
 (x_{i}, y_{j},z_{k}) ,
    \, i=0, \ldots , N_x+1 ,
    \, j=0, \ldots , N_y+1 ,
    \, k=0 , \ldots , N_z+1 \} .
$$

We introduce then the staggered grids along all coordinates
\begin{eqnarray*}
%\label{tsyb-staggered}
 x_{i+1/2} = \frac12 (x_i + x_{i+1}) , \quad i=0, \ldots , N_x ,
 \\
 \nonumber
 y_{j+1/2} = \frac12 (y_j + y_{j+1}) , \quad j=0, \ldots , N_y ,
 \\
 \nonumber
 z_{k+1/2} = \frac12 (z_k + z_{k+1}) , \quad k=0, \ldots , N_z  .
\end{eqnarray*}

%Here $h_{i+1/2}$, $h_i$, $g_{j+1/2}$, and $g_j$ are
%the step sizes.

The velocities $v^1$, $v^2$ and $v^3$  are defined on the  grids which are staggered respectively along the corresponding coordinates
\begin{eqnarray*}
   \omega_{1} = \{ (x_{i}, y_{j+1/2}, z_{k+1/2}),
    \, i=0, \ldots , N_x+1,
    \, j=0 , \ldots , N_y,
    \, k=0 , \ldots , N_z \} ,
\label{omega_v1}
\\
   \omega_{2} = \{ (x_{i+1/2}, y_{j}, z_{k+1/2}),
    \, i=0, \ldots , N_x,
    \, j=0, \ldots , N_y+1,
    \, k=0 , \ldots , N_z \} ,
\label{omega_v2}
\\
   \omega_{3} = \{ (x_{i+1/2}, y_{j+1/2}, z_{k}),
    \, i=0, \ldots , N_x ,
    \, j=0, \ldots , N_y ,
    \, k=0 , \ldots , N_z+1 \} .
\label{omega_v3}
\end{eqnarray*}
Finally, the pressure $p$ is defined at the nodes
$$
 \omega_p = \{
 (x_{i+1/2}, y_{j+1/2},z_{k+1/2}) ,
    \, i=0, \ldots , N_x ,
    \, j=0, \ldots , N_y ,
    \, k=0 , \ldots , N_z \} .
$$
In the case of Dirichlet boundary conditions  we do not  need any fictitious nodes for the discretization.
In the case of mixed boundary conditions the grids are introduced in such a way that on $\partial_2 D$ the boundary conditions for the temperature and the normal component of velocity are fulfilled automatically.
We define  fictitious nodes for the temperature and velocity  $v^2$ to approximate the boundary conditions on $\partial_1 D$.

\subsection{Discrete finite-difference operators}
To approximate (\ref{eq_vec})--(\ref{ic}) we define a set of discrete analogs of first order differential operators  on a two-point stencil
\begin{eqnarray}
\nonumber
&&
d_1\theta_{i+1/2,j,k}
= \frac {\theta_{i+1,j,k}-\theta_{i,j,k}} { x_{i+1}-x_j}
\approx (\theta_x)_{i+1/2,j,k},
\\
\label{Dif_Op}
&&
d_2 \theta_{i,j+1/2,k}
=\frac{\theta_{i,j+1,k}-\theta_{i,j,k}} { y_{j+1}-y_j}
\approx (\theta_y)_{i,j+1/2,k},
\\
\nonumber
&&
d_3 \theta_{i,j,k+1/2}
=\frac{\theta_{i,j,k+1}-\theta_{i,j,k}} { z_{k+1}-z_k}
\approx (\theta_z)_{i,j,k+1/2},
\end{eqnarray}
and weighted averaging operators on the coordinates
\begin{eqnarray}
\nonumber
&&
\delta_{1} \theta_{i+1/2,j,k} =
\frac{ (x_{i+1} - x_{i+1/2}) \theta_{i+1,j,k}
      + (x_{i+1/2} - x_i) \theta_{i,j,k}
     }
     { x_{i+1} - x_i }
\approx (\theta)_{i+1/2,j,k},
\\
\label{Ave_Op}
&&
 \delta_{2} \theta_{i,j+1/2,k} =
\frac{ (y_{j+1} - y_{j+1/2}) \theta_{i,j+1,k}
      + (y_{j+1/2} - y_j) \theta_{i,j,k}
     }
     { y_{j+1} - y_j }
\approx (\theta)_{i,j+1/2,k},
\\
\nonumber
&&
 \delta_{3} \theta_{i,j,k+1/2} =
\frac{ (z_{k+1} - z_{k+1/2}) \theta_{i,j,k+1}
      + (z_{k+1/2} - z_k) \theta_{i,j,k}
     }
     { z_{k+1} - z_k }
\approx (\theta)_{i,j,k+1/2} .
\end{eqnarray}
The formulas (\ref{Dif_Op})--(\ref{Ave_Op}) are valid both for integer
and half-integer values of $i$, $j$ and $k$.
Then the discrete analog of the Laplacian on the seven-nodes stencil can be written as
\begin{equation}
\triangle_h = d_1 d_1 + d_2 d_2 + d_3 d_3 \approx \triangle  ,
\end{equation}
and the averaging operator on three-dimensional cell is given as \begin{equation}
\delta_0 = \delta_1 \delta_2 \delta_3 .
\end{equation}

The nonlinear term approximation is constructed using a linear combination of two terms
\begin{eqnarray}
\label{Appr_nonlin}
(v \cdot\nabla \theta )_{i,j,k}
&\approx& J(\theta,v)_{i,j,k} =
\\
\nonumber
&=& \frac{1}{3}
\left[
d_1 \delta_{1} (\theta \delta_{2} \delta_{3} v^1)
+ d_2 \delta_{2} (\theta \delta_{1} \delta_{3} v^2)
+ d_3 \delta_{3} (\theta \delta_{1} \delta_{2} v^3)
\right]_{i,j,k}
\\
\nonumber
&+& \frac{2}{3}
\left[
d_1\delta_{2}\delta_{3} (\delta_0 \theta \delta_{1} v^1)
+ d_2\delta_{1}\delta_{3} (\delta_0 \theta \delta_{2} v^2)
+ d_3\delta_{1}\delta_{2}(\delta_0 \theta \delta_{3} v^3)
\right]_{i,j,k}.
\end{eqnarray}
%NEW
This provides second order accuracy for the uniform grid and an
asymptotically second order accuracy for the quasi-uniform mesh.
It allows conservation of energy and constitute a mimetic
discretization of the underlying problem.
%NEW

%It was shown in \cite{MoiFry80} that this formulae allows to conserve energy and total or mean square vorticity (enstrophy) in the finite-difference approximation of two-dimensional flow and thus being mimetic for the underlying problem.
%

\subsection{Semi-discretization}
To reach some steady state it is useful to apply an approach based
on artificial compressibility \cite{Ch67,Yanenko66}. Thus instead of
equation (\ref{eq_con}) we consider the following equation with
coefficient $\eta$
\begin{equation}
  \dot p + \frac{1}{\eta} \nabla\cdot v= 0 .
\label{ArtComp}
\end{equation}

Using the operators (\ref{Dif_Op})--(\ref{Appr_nonlin}) we discretize  system (\ref{eq_vec}), (\ref{eq_tem}) and (\ref{ArtComp}) in the following form
\begin{eqnarray}
\label{DdT}
&&
\left[
\dot{\theta} - \triangle_h \theta - \lambda \delta_1 \delta_2 v^{3}
+ J(\theta,v)
\right]_{i,j,k} = 0 ,
\\
\label{DdV1}
&&
\left[ \dot{v^1} - \varepsilon (- d_1 p - v^{1})
\right]_{i,j+1/2,k+1/2} = 0 ,
\\
\label{DdV2}
&&
\left[ \dot{v^2} - \varepsilon (- d_2 p - v^{2})
\right]_{i+1/2,j,k+1/2} = 0 ,
\\
\label{DdV3}
&&
\left[ \dot{v^3} -
\varepsilon (- d_3 p - v^{3} + \delta_1 \delta_2 \theta )
\right]_{i+1/2,j+1/2,k} = 0 ,
\\
\label{DdP}
&&
\left[\dot{p} + \frac{1}{\eta} ( d_1 v^{1}+ d_2 v^{2}+d_3 v^{3})
\right]_{i+1/2,j+1/2,k+1/2}  = 0 .
\end{eqnarray}

For the problem with Dirichlet boundary conditions  (\ref{eq_vec})--(\ref{bc_A}) the grids are introduced in such a way that the nodes for the transversal velocity are located on the wall.
It allows us to fulfill the boundary conditions on a rigid wall (\ref{bc_A}) in a very simple way for each pair of planes:
\\
for $x=0$ ($i=0$) and $x=L_x$ ($i=N_x+1$) we have
\begin{eqnarray}
\label{d_bc_1}
&&
v^1_{i,j+1/2,k+1/2} = 0 ,
 \quad j=0, \ldots ,N_y ,
 \quad k=0, \ldots ,N_z ,
\\
\nonumber
&& \theta_{i,j,k} = 0 ,
  \quad j=0, \ldots ,N_y+1, \quad k=0, \ldots ,N_z+1 ,
\end{eqnarray}
for $y=0$ ($j=0$) and $y=L_y$ ($j=N_y+1$ )we have
\begin{eqnarray}
\label{d_bc_2}
&&
v^2_{i+1/2,j,k+1/2} = 0 ,
 \quad i=0,  \ldots ,N_x ,
 \quad k=0, \ldots ,N_z ,
\\
\nonumber
&& \theta_{i,j,k} = 0 ,
 \quad i=0, \ldots ,N_x+1,  \quad k=0, \ldots ,N_z+1 ,
\end{eqnarray}
and for $z=0$ ($k=0$) and $z=L_z$ ($k=N_z+1$) we have
\begin{eqnarray}
\label{d_bc_3}
&&
v^3_{i+1/2,j+1/2,k} = 0 ,
 \quad i=0, \ldots ,N_x ,
 \quad j=0, \ldots ,N_y ,
\\
\nonumber
&&
\theta_{i,j,k} = 0 ,
\quad  i=0, \ldots ,N_x+1, \quad j=0, \ldots ,N_y+1
.
%\label{bc_tem}
\end{eqnarray}

The problem with mixed boundary conditions  (\ref{eq_vec})--(\ref{bc_vel}) and (\ref{bc_B}) is discretized using  fictitious nodes to satisfy the boundary conditions on the planes $y=0$ and $y=L_y$.
Thus instead of (\ref{d_bc_2}) we have
\begin{eqnarray}
\label{d_bc_2mix}
&&
v^2_{i+1/2,0,k+1/2} = - v^2_{i+1/2,1,k+1/2} ,
 \quad i=0, \ldots ,N_x ,
 \quad k=0, \ldots ,N_z ,
\\
\nonumber
&&
v^2_{i+1/2,N_y+1,k+1/2} = - v^2_{i+1/2,N_y,k+1/2} ,
 \quad i=0, \ldots ,N_x ,
 \quad k=0, \ldots ,N_z ,
\\
&&
\theta_{i,0,k} = \theta_{i,1,k} ,
\quad \theta_{i,N_y+1,k} = \theta_{i,N_y,k} ,
 \quad i=0, \ldots ,N_x+1 ,
 \quad k=0, \ldots ,N_z+1 .   \nonumber
\end{eqnarray}

\subsection{Discrete equations for the planar problem}
The system of ordinary differential equations (\ref{DdT})--(\ref{DdP}), (\ref{d_bc_1}), (\ref{d_bc_3}) and (\ref{d_bc_2mix}) has a solution such that $v^2_{i+1/2,j,k+1/2}=0$ and the other variables are invariant w.r.t. index $j$.
Then, we can exclude equation (\ref{DdV2}) and consider the two-dimensional problem.
Moreover, we can eliminate  pressure and velocities and reformulate the system with the discrete stream function and temperature.
Let's define the stream function $\psi$ at the nodes $\omega_{0}$  using difference operators (\ref{Dif_Op})
\begin{equation}
\label{dpsi}
(d_3 \psi)_{i,k+1/2} = - v^1_{i,k+1/2}, \quad
(d_1 \psi)_{i+1/2,k} = v^3_{i+1/2,k}.
\end{equation}
The discrete analog of the continuity equation (\ref{eq_con}) is
automatically fulfilled by (\ref{dpsi}). After the substitution of
(\ref{dpsi}) in (\ref{DdV1}) and (\ref{DdV3}) and the combination of
the resulting formulas we can deduce the analog of (\ref{eq_psi})
(inertia terms are omitted)
\begin{equation}
\label{Delta_psi}
\left[ \Delta_{2,h} \psi - D_x \theta \right]_{i,k} = 0.
\end{equation}
Similarly  we find from (\ref{DdT})
\begin{equation}
\label{Delta_theta}
\dot \theta_{i,k} = \left[ \Delta_{2,h} \theta
+ \lambda D_x\psi + \frac13 J_D + \frac23 J_d  \right]_{i,k} .
\end{equation}
Here  $D_x=d_1 \delta_1$, $D_z=d_3 \delta_3$ are the first order
differencing operators on  three-nodes stencils.
The Laplacian and Jacobian are approximated using
\begin{eqnarray}
\label{jac}
&& \Delta_{2,h}  =  d_1 d_1 + d_3 d_3,
\nonumber \\
&& J_D  =
   D_x \left( \theta D_y \psi \right)
 - D_y \left( \theta D_x \psi \right),
 \\
 && J_d  =
 \widehat{d_x} \left( \widehat{d_0} \theta \widehat{d_z} \psi \right)
     - \widehat{d_z} \left( \widehat{d_0} \theta \widehat{d_x} \psi \right)
\nonumber ,
\end{eqnarray}
where $\widehat{d_0}=\delta_1 \delta_3$, $\widehat{d_x}=d_1 \delta_3$, $\widehat{d_z}=d_3 \delta_1$.

The resulting scheme (\ref{Delta_psi})--(\ref{jac}) ensures the
fulfillment of a discrete analog  of the cosymmetry property
(\ref{int}) as well as the nullification of the gyroscopic terms
\cite{KarTsy05a}. Equations in \cite{KarTsy99} follow from
(\ref{Delta_psi})--(\ref{jac}) for the case of the uniform grid $x_i
= i h$, $z_k  = k g$, $h = L_x/(N_x+1)$,  $g=L_z/(N_z+1)$. Then the
Jacobian approximation gives the famous Arakawa formula \cite{Ara66}
on uniform grids $h=g$.

\section{Computation of the family of steady states}
We rewrite the resulting system of equations in vector form. Let us
introduce   vectors which contain only unknowns at internal nodes
\begin{eqnarray*}
&&
\Theta = (\theta_{1,1,1},\ldots, \theta_{N_x,1,1}, \theta_{1,2,1},\ldots,
\theta_{N_x, N_y, N_z}) ,
\\
&&
V^1 = (v^1_{1,1/2,1/2},\ldots,
v^1_{N_x,1/2,1/2}, v^1_{1,3/2,1/2}, \ldots,
v^1_{N_x,N_y + 1/2,N_z + 1/2}) ,
\\
&&
V^2 = (v^2_{1/2,1,1/2},\ldots,
v^2_{N_x+1/2,1,1/2}, v^2_{1/2,2,1/2}, \ldots,
v^2_{N_x + 1/2,N_y,N_z + 1/2}) ,
\\
&&
V^3 = (v^3_{1/2,1/2,1},\ldots,
v^3_{N_x+1/2,1/2,1}, v^3_{1/2,3/2,1}, \ldots,
v^3_{N_x + 1/2,N_y + 1/2, N_z}) ,
\\
&&
P = (p_{1/2,1/2,1/2},\ldots,
p_{N_x+1/2,1/2,1/2}, p_{1/2,3/2,1/2}, \ldots,
p_{N_x + 1/2,N_y + 1/2,N_z + 1/2}),
\end{eqnarray*}
and obtain the system which corresponds (\ref{DdT})--(\ref{DdP})
\begin{eqnarray}
\label{ATeqRCv}
&&
\dot{\Theta} = A_1 \Theta + \lambda C_1 V^3 - F(\Theta,V), \nonumber
\\
\nonumber
&&
 \dot{V^1} = - B_4 P - C_2 V^1,
\\
\nonumber
&&
 \dot{V^2} = - B_5 P - C_3 V^2,
\\
 &&
\dot{V^3} = - B_6 P - C_4 V^3 + C_5 \Theta,
\\
\nonumber
&&
\dot{P} = - B_1 V^1 - B_2 V^2 - B_3 V^3.
\end{eqnarray}
Here the matrices $B_k$, $k=1, \ldots , 6$, are constructed by the application of first order difference operators, and the matrices  $C_k$, $k=1, \ldots , 5$,   by  the averaging operators.
The matrix $A_1$ presents the discrete form of the Laplacian.
The nonlinear term is given by $F(\Theta,V)$.
Equations (\ref{ATeqRCv}) form a system of
$$5N_x N_y N_z+3(N_x N_y+N_x N_z+N_y + N_z)+2(N_x+ N_y+ N_z)+1$$ unknowns.

From (\ref{ATeqRCv}) at $J = 0$ we can derive the perturbation
equations ($\sigma$ is a decrement of linear growth) to analyze the
stability of the state of rest
\begin{eqnarray}
\label{EqT}
&& \sigma\Theta = A_1 \Theta + \lambda C_1 V^3,
\\
\label{Eq1}
&&  \sigma V^1 = - B_4 P - C_2 V^1,
\\
\label{Eq2}
&&  \sigma V^2 = - B_5 P - C_3 V^2,
\\
\label{Eq3}
&& \sigma V^3 = - B_6 P - C_4 V^3 + C_5 \Theta,
\\
\label{EqP}
&&  \sigma P = - B_1 V^1 - B_2 V^2 - B_3 V^3.
\end{eqnarray}

For the decrement $\sigma=0$ we obtain the system from which we can determine the threshold value of the Rayleigh number corresponding to the monotonic loss of stability.
We can express $P$, $V^1$, $V^2$ и $V^3$ via $\Theta$ from (\ref{EqT})--(\ref{EqP}) and obtain a system of $N_x N_y  N_z$  equations for the unkown vector $\Theta$.
Substituting (\ref{Eq1}), (\ref{Eq2}) and (\ref{Eq3}) into (\ref{EqP}) we
deduce
\begin{equation}
A_1 \Theta = \lambda C_1 (C_5 - B_6 S) \Theta .
\label{Part_Lin_Sys}
\end{equation}
Here we find the vector $P=S\Theta$ from the system of linear algebraic equations with rank deficiency 1
$$
(B_1 C_2^{-1} B_4 + B_2 C_3^{-1} B_5 + B_3 C_4^{-1} B_6) P = B_3 C_4^{-1} C_5 \Theta.
$$
Since for an incompressible flow the pressure may differ by a constant we can exclude one component of $P$ and respectively one equation.

%To find the convective steady patterns we apply the direct approach, i.e., the system of ordinary differential equations (\ref{ATeqRCv})
%is integrated by the classical fourth order Runge-Kutta method.
%NEW
To find an isolated convective pattern we apply the direct approach and integrate the system of ordinary differential equations (\ref{ATeqRCv}) by the classical fourth order Runge-Kutta method up to convergence.

To compute a family of steady states we apply the technique based on
the cosymmetric version of the implicit function theorem
\cite{Yud96} and the algorithm developed in
\cite{Gov98,KarTsy99,Gov00}.
The zero equilibrium $V^1=V^2=V^3=P=\Theta=0$ is globally stable for
$\lambda<\lambda_{1}$ where $\lambda_1$ is the minimal eigenvalue for spectral problem
(\ref{Part_Lin_Sys}).
When $\lambda$ is slightly larger than
$\lambda_{1}$,  then all points of the family are stable
\cite{Yud95}. Starting from the vicinity of unstable zero
equilibrium we integrate the ordinary differential equations
(\ref{ATeqRCv}) up to a point $\Theta_0$ close to some stable
equilibrium on the family. Then we correct the point $\Theta_0$ by
the Newton method. To predict the next point on the family we
determine the kernel of the linearization matrix (Jacobi matrix) at
the point $\Theta_0$ and then use the Adams-Bashford method. This
procedure is repeated to obtain the entire family of steady states.
It is important to note that the given procedure allows us to
compute the stable regimes as well as unstable ones.

\section{Numerical results}
V. Yudovich \cite{Yud91} proved that the appearance of a family of
steady states in the planar Darcy convection is caused by the
nontrivial cosymmetry of the problem. Loss of stability of the state
of rest is characterized by the repeated eigenvalues for the
corresponding spectral problem. For the problem under consideration
we find the critical Rayleigh numbers from the system
(\ref{Part_Lin_Sys}). There exist two scenarios of instability of
the state of rest in the parallelepiped: branching off of isolated
regimes and the appearance of a family of steady states
\cite{BLT98}. In our computer experiment the emergence of the family
was only observed for the problem with mixed boundary conditions. It
was found that a family of stable equilibria has appeared in the
case of rather small value of  $L_y/L_x$ (relative depth) which
depends also on the ratio $L_x/L_z$.

It was shown  in \cite{KarTsy05} that a uniform grid is the best
choice for the computation of the critical Rayleigh numbers while a
nonuniform grid is useful for the  computation of a specific regime
with a desirable accuracy. Thus, we use the uniform grids and set up
the amount of internal nodes for the temperature as $N_x \times N_y
\times N_z$.

%\begin{table}[h]
%\caption{Dependence of the critical Rayleigh numbers on the depth $L_y$ for two problems: Dirichlet b.c. (mixed b.c.);
%$L_x = 2$, $L_z=1$, mesh $24 \times 6 \times 12$  }
%\label{Different_Grids}
%\begin{center}
%\begin{tabular}{c|c|c|c|c|c}
% $L_y $ &  0.4 &  0.6 &  0.8 &  1.2 & 2 \\
%\hline
%$\lambda_{1}$ & (50,5) & (50,5) & (46,8) & - & - \\
%$\lambda_{2}$ & (50,5) & (50,5) & (50,5) & -& - \\
%$\lambda_{3}$ & (83,7) & (56,5) & (50,5) & - & - \\
%$\lambda_{4}$ & (83,7) & (65,3) & (55,0) & - & - \\
%$\lambda_{5}$ & (90,4) & (80,1) & (69,3) & - & - \\
%$\lambda_{6}$ & (99,6) & (83,7) & (83,7) & - & - \\
%$\lambda_{7}$ & (115,3) & (83,7) & (83,7) & - & - \\
%\end{tabular}
%\end{center}
%\end{table}
\subsection{Critical Rayleigh numbers}
We present in Table~\ref{Crit_Dirichlet}  the first seven critical Rayleigh numbers $\lambda$ for the problem with  Dirichlet boundary conditions for several values of the depth  $L_y$ and fixed length $L_x = 2$ and height $L_z=1$.
In the case $L_y<1$ ($L_y>1$) the computation of critical Rayleigh numbers was carried out on the mesh of  $14 \times 6 \times 6$ ($14 \times 10 \times 6$)  internal nodes for the temperature.
\begin{table}[h]
\caption{Dependence of critical Rayleigh numbers on the depth $L_y$ for the problem with Dirichlet boundary conditions;
$L_x = 2$, $L_z=1$, mesh $14 \times 6 \times 6$ ($14 \times 10 \times 6$) }
\label{Crit_Dirichlet}
\begin{center}
\begin{tabular}{c|c|c|c|c|c}
 $L_y $       &  0.4  &  0.6  &  0.8 &  1.2        & 1.6 \\     % 2.0
%                                                               (12*12*6)
\hline
$\lambda_{1}$ & 157.5 & 99.2  & 77.9 & 62.0 (59.1) & (53.5) \\  % 51.3
$\lambda_{2}$ & 159.0 & 100.1 & 78.0 & 62.3 (59.7) & (54.9) \\  % 53.8
$\lambda_{3}$ & 207.0 & 138.4 & 108.5 & 83.4 (73.3) & (60.6) \\  % 53.8
$\lambda_{4}$ & 209.4 & 139.6 & 116.0 & 85.1 (78.2) & (66.0) \\  % 59.5
$\lambda_{5}$ & 298.3 & 178.7 & 122.9 & 90.0 (79.6) & (69.5) \\  % 63.9
$\lambda_{6}$ & 303.8 & 191.1 & 132.4 & 100.1 (95.2) & (81.4) \\ % 75.2
$\lambda_{7}$ & 310.5 & 192.0 & 152.1 & 107.6 (95.3) & (86.9) \\ % 75.2
\end{tabular}
\end{center}
\end{table}

One can see that some critical values in Table~\ref{Crit_Dirichlet}
are close to each other. When we take $L_x=L_y$ we find that some
critical values coincide. For example, for $D=[0,2] \times [0,2] \times [0,1]$
and the mesh $12 \times 12 \times 6$ we find that $\lambda_1=51.3$
and $\lambda_2=\lambda_3=53.8$.
%But this is a sequence of a symmetry of the problem and doesn't lead to the appearance of a family of steady states.
%NEW
But this is a consequence of the discrete symmetries on the $x_1$
and $x_2$ coordinates and doesn't lead to the appearance of a family
of steady states.

We summarize in Table~\ref{Crit_mixed_bc} the  first seven critical
Rayleigh numbers $\lambda$ for the  mixed boundary conditions
problem  for several values of the depth  $L_y$ and fixed length
$L_x = 2$ and height $L_z=1$. One can see that for  $L_y = 0.4$ and
$ L_y = 0.6$ two minimal eigenvalues of the problem
(\ref{Part_Lin_Sys}) are  repeated. This corresponds to the  birth
of the continuous family of steady states. The parallelepiped with
$L_y = 0.8$ gives a single minimal eigenvalue and that results in
the appearance of two isolated stationary regimes.
\newpage
\begin{table}[t]
\caption{Dependence of critical Rayleigh numbers on the depth $L_y$ for the problem with mixed boundary conditions;
$L_x = 2$, $L_z=1$, mesh $14 \times 6 \times 6$ ($14 \times 10 \times 6$) }
\label{Crit_mixed_bc}
\begin{center}
\begin{tabular}{c|c|c|c|c|c}
 $L_y $ &  0.4 &  0.6 &  0.8 &  1.2 & 1.6 \\    % 2.0
\hline
$\lambda_{1}$ & 52.5 & 52.5 & 46.7 & 44.5 (42.5) & (45.7) \\  % 45.5
$\lambda_{2}$ & 52.5 & 52.5 & 52.5 & 51.6 (49.4) & (46.1) \\  % 48.7
$\lambda_{3}$ & 90.8 & 56.6 & 52.5 & 52.5 (52.5) & (49.6) \\  % 50.5
$\lambda_{4}$ & 93.1 & 66.8 & 56.2 & 52.5 (52.5) & (52.5) \\  % 52.5
$\lambda_{5}$ & 93.1 & 84.7 & 73.2 & 67.8 (58.1) & (52.5) \\  % 52.5
$\lambda_{6}$ & 102.1 & 93.1 & 93.1 & 68.3 (64.8) & (57.8) \\ % 54.2
$\lambda_{7}$ & 122.2 & 93.1 & 93.1 & 80.9 (68.6) & (66.3) \\ % 60.0
\end{tabular}
\end{center}
\end{table}

It should be noted that on the fixed mesh the threshold corresponding to the branching off of the family of steady states (repeated critical values) doesn't depend on the depth $L_y$.
Comparison with results on the mesh $24 \times 6 \times 12$ shows that even a rough mesh allows to find a threshold (minimum of critical Rayleigh numbers) with accuracy about 10\%.
For instance, using the mesh $24 \times 6 \times 12$ we obtain $\lambda_1=50.5$ for $L_y=0.4$ and $L_y=0.6$ and $\lambda_1=46.8$ for $L_y=0.8$.

On the other hand convection in a three-dimensional box with
insulating impermeable lateral boundaries (the lateral walls are
taken as thermally insulating) was a subject of  numerous works
\cite{NiBe99}. We may refer here to \cite{StrSch79} where it was
shown that different two-dimensional and three-dimensional states
appear depending on the initial conditions.

\subsection{Computation of steady states}
Now we study the mixed boundary conditions problem and consider the
parallelepipeds with  small depth when the family of stationary
solutions branches off. The steady states belonging to the family
are essentially two-dimensional and don't depend on the coordinate
$y$.

\begin{figure} [h]
\centering
\includegraphics[width=13.0cm]{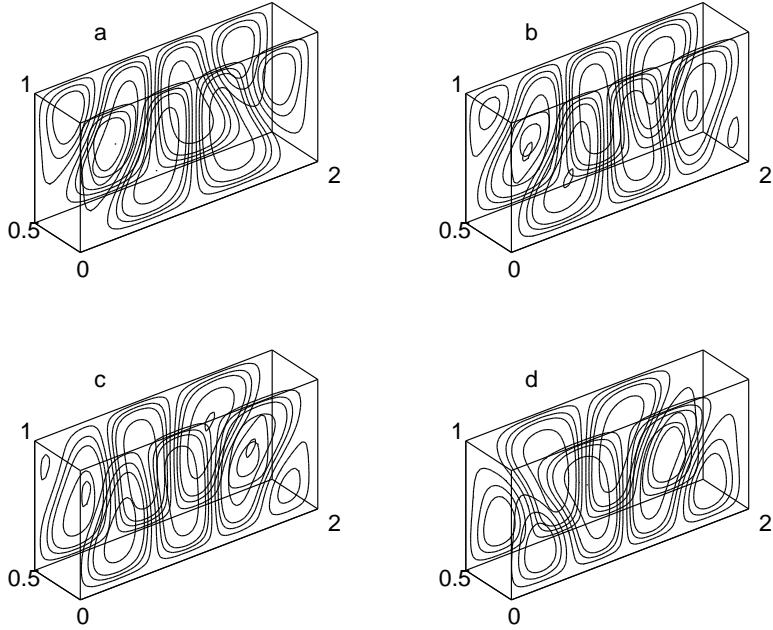}
\caption{Members of the family of steady states;  $\lambda = 60$,
$D = [0,2] \times [0,0.5] \times [0,1]$}
\label{R200_stream3_mix}
\end{figure}

\begin{figure} [h]
\centering
\includegraphics[width=10.0cm]{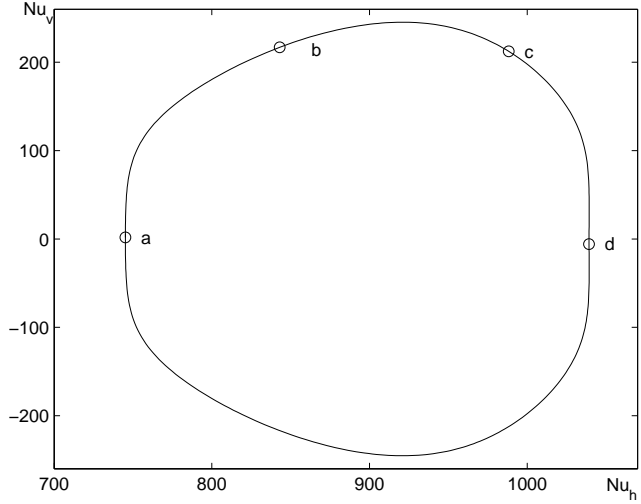}
\caption{Family of steady states;  $\lambda = 60$, $D = [0,2] \times [0,0.5] \times [0,1]$}
\label{fam_2x1_R200}
\end{figure}

Fig.~\ref{R200_stream3_mix} demonstrates this by the presentation of several convective patterns from the family: none has transversal motion to the plane $y=const$.
The relative location of each steady state is given in Fig.~\ref{fam_2x1_R200} where each letter corresponds to a flow pattern in Fig.~\ref{R200_stream3_mix}.
In Fig.~\ref{fam_2x1_R200} we use the Nusselt values for the planar problem \cite{Gov98}
\begin{equation}
\label{Nu}
Nu_v  =  \int_0^{L_z} \theta_x(\frac{L_x}{2},0,z) dz,
\quad
Nu_h  =  \int_0^{L_x} \theta_z(x,0,0) dx .
\end{equation}
Here $Nu_v$ corresponds to the cumulative heat flux from  left to right defined for the centered vertical section of the rectangular domain.
The value $Nu_h$ is a combined heat flux through the bottom of the enclosure, $y=0$.

\begin{figure} [h]
\centering
\includegraphics[scale=0.55, angle=-90]{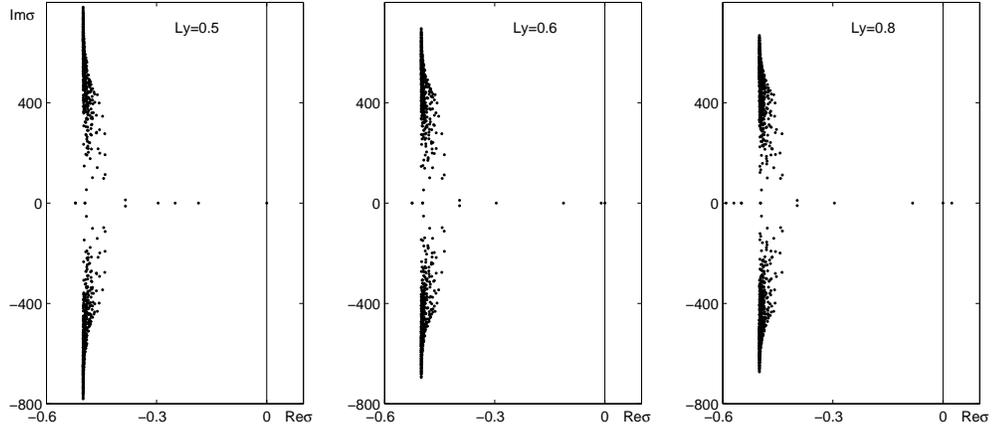}
\caption{Stability spectrum for the two-rolls steady state from the family;  $\lambda = 100$, $L_x = 2$, $L_z=1$}
\label{spectr_Ly_R100_sym}
\end{figure}

The regimes in Fig. 2 are the members of the family of steady states,  because  stability spectra for each state have exactly one value being zero with reliable accuracy ($10^{-8}$).
We plot the distribution of eigenvalues of the Jacobi matrix (matrix of linearization) computed for the convective pattern with two symmetrical rolls (regime a in Fig.~\ref{fam_2x1_R200}).
 We take different values of the depth $L_y$ to demonstrate the three-dimensional instability on the family, see Fig.~\ref{spectr_Ly_R100_sym}).
It is clearly seen that exactly one point $\sigma$ lies on the imaginary axis.
The corresponding eigenvector defines  the neutral direction along the family.
Such a family is called cosymmetric, the stability of its members is governed by nonzero eigenvalues or  defined on the submanifold being transversal to the family.

\begin{figure} [h]
\centering
\includegraphics[scale=0.55, angle=-0]{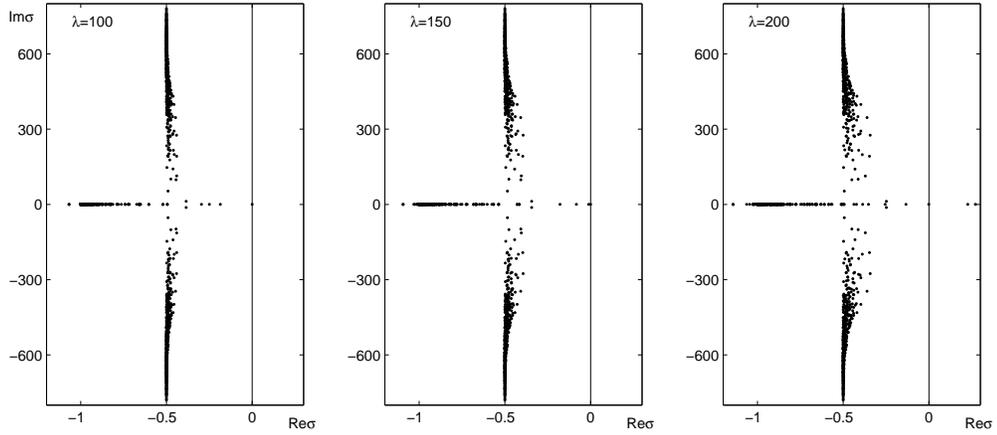}
\caption{Stability spectrum for symmetric two-rolls pattern from the family for different $\lambda$; $D = [0,2] \times [0,0.5] \times [0,1]$}
\label{spectr_Ly05_R100_200_sym}
\end{figure}

It should be noted that  the given Rayleigh number is rather far from the critical value when some members of the family become unstable.
In the case of the parallelepiped $D = [0,2] \times [0,0.5] \times [0,1]$ the family of steady states appears at $\lambda \approx 51$.
The transformation of stability spectra  with increasing  $\lambda$ is presented in Fig.~\ref{spectr_Ly05_R100_200_sym}.
One can see that the symmetrical convective pattern lost its stability before $\lambda=200$.
Some states on the family become unstable at $\lambda \approx 190$.
This value is less than  $\lambda \approx 400$ which corresponds to the critical value of instability for the planar problem.

\begin{figure} [h]
\centering
\includegraphics[scale=0.55, angle=-90]{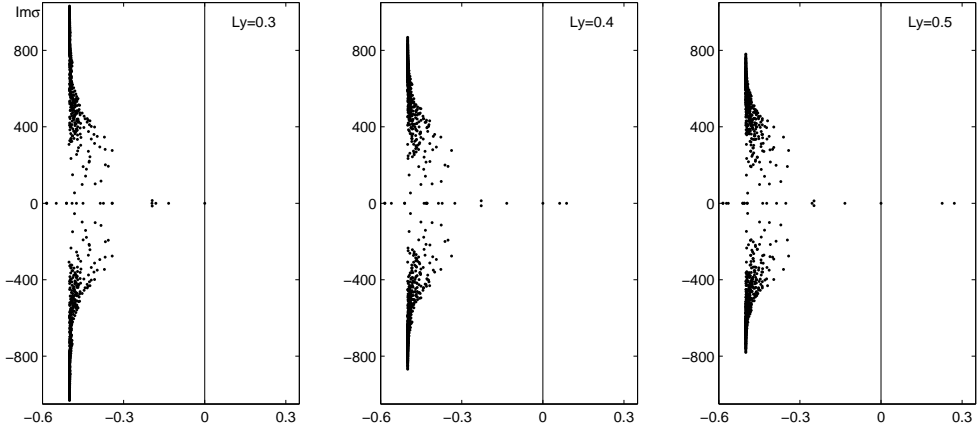}
\caption{Stability spectrum for the two-rolls steady state from the family;  $\lambda = 200$, $L_x = 2$, $L_z=1$}
\label{spectr_Ly_R200_sym}
\end{figure}

We present  in Fig.~\ref{spectr_Ly_R200_sym}  the distribution of
spectra for different values of the depth $L_y$. One can see that
the steady state with two rolls is stable when $L_y=0.3$ and
unstable for $L_y \ge 0.4$ (two spectra values in right part of the
complex plane). When the Rayleigh number becomes greater the
instability on the family occurs for smaller values of the depth
$L_y$.

\begin{figure} [h]
\centering
\includegraphics[width=13.0cm]{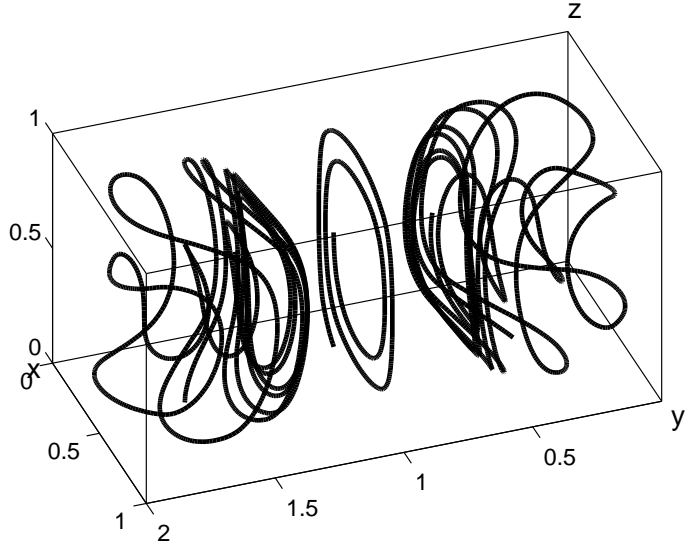}
\caption{Isolated steady state;  $\lambda = 52$, $D = [0,2] \times [0,0.9] \times [0,1]$}
\label{picEq_isolyr_nonsym}
\end{figure}

\begin{figure} [h]
\centering
\includegraphics[width=11.0cm]{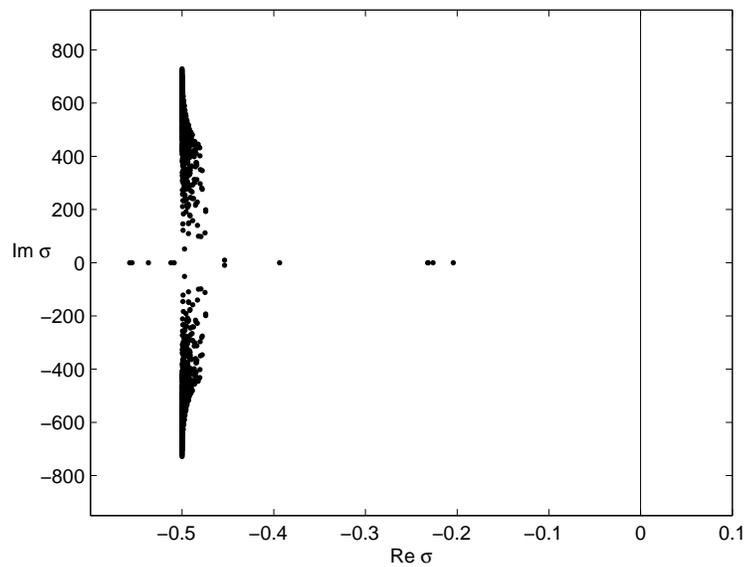}
\caption{Stability spectrum for the isolated steady state from the family;  $\lambda = 52$, $D = [0,2] \times [0,0.9] \times [0,1]$}
\label{spectr_2x09x1_12x6x6_r52_nonsym}
\end{figure}

When the depth $L_y=0.9$  the only isolated stable regimes are branched off of the state of rest.
Because of the discrete symmetries of the problem  (\ref{sym_x})--(\ref{sym_z}) we obtained two regimes.
The flow pattern for one of these regimes is presented in Fig.~\ref{picEq_isolyr_nonsym}.
The distribution of eigenvalues for this  stable steady state is displayed in Fig.~\ref{spectr_2x09x1_12x6x6_r52_nonsym}.
One can see that the distribution of the stability spectra has no point  close to the imaginary axis.
This regime is essentially three-dimensional and isolated.
Branching off the isolated convective regimes is characteristic for the parallelepipeds with non-small depth.

\section{Conclusion}
Convection in a porous  parallelepiped has demonstrated two different scenarios of instability of the state of rest.
It was found  by  computer experiments that for zero heat fluxes on two lateral sides (mixed boundary conditions)  the appearance of a cosymmetric continuous family of equilibria becomes possible.

To compute a continuous family of equilibria one needs to solve repeatedly a nonlinear system of algebraic equations that is degenerated in the vicinity of the family.
This is why discretization is so important for the  Darcy convection.
We have developed the approach based on primitive variables equations and a finite-difference discretization with staggered nonuniform grids.
This scheme mimics the nontrivial characteristics of the underlying problem that admits  existence of a continuous family of steady states.

\subsection*{Acknowledgements}
The authors thank to the referee for careful reading and advises. The authors acknowledge the support of Guest Researcher Fellowship Programme   of T\"{U}BITAK (Turkish Scientific Research Council).
A.N. and V.T. were partially supported by the Program for the leading scientific schools (\# 5747.2006.1)
and Russian Foundation for Basic Research (\# 08-01-00734) and the
Program for National Universities of Russia.

%\vspace*{1cm}

\bibliographystyle{amsalpha}

\end{document}